\def\Msun{\hbox{M$_\odot$}}

\def\Zsol{\hbox{Z$_\odot$}}

\def\kms{\hbox{km$\,$s$^{-1}$}}
\def\cm3{\hbox{cm$^{-3}$}}
\def\cm2{\hbox{cm$^{-2}$}}
\def\s1{\hbox{s$^{-1}$}}

\def\one{\,{\sc i}}             
\def\two{\,{\sc ii}}
\def\three{\,{\sc iii}}
\def\four{\,{\sc iv}}
\def\five{\,{\sc v}}
\def\six{\,\textsc{vi}}

\def\ha{H$\alpha$}

\def\mka{Mrk~71-A}
\def\mkb{Mrk~71-B}

\documentclass[twocolumn]{aastex631}

\usepackage{amsmath}
\usepackage{amssymb}
\usepackage{mathrsfs}
\usepackage{upgreek}
\usepackage{graphicx}
\usepackage{overpic}
\usepackage{threeparttable}

\received{XXX}
\revised{XXX}
\accepted{\today}
\submitjournal{ApJ}

\shorttitle{HST FUV spectroscopy of SSC A in Mrk 71}
\shortauthors{L.\ J.\ Smith et al.}

\begin{document}


\title{HST FUV Spectroscopy of Super Star Cluster A in the Green Pea Analog Mrk 71: Revealing the Presence of Very Massive Stars}
\correspondingauthor{Linda\ J.\ Smith}
\email{lsmith@stsci.edu}
\author[0000-0002-0806-168X]{Linda J. Smith}
\affiliation{Space Telescope Science Institute, 3700 San Martin Drive, Baltimore, MD 21218, USA}

\author[0000-0002-5808-1320]{M.S. Oey}
\affiliation{Astronomy Department, University of Michigan, Ann Arbor, MI 48103, USA}

\author[0000-0003-4857-8699]{Svea Hernandez}
\affiliation{AURA for ESA, Space Telescope Science Institute, 3700 San Martin Drive, Baltimore, MD 21218, USA}

\author[0000-0002-2918-7417]{Jenna Ryon}
\affiliation{Space Telescope Science Institute, 3700 San Martin Drive, Baltimore, MD 21218, USA}

\author[0000-0003-2685-4488]{Claus Leitherer}
\affiliation{Space Telescope Science Institute, 3700 San Martin Drive, Baltimore, MD 21218, USA}

\author[0000-0003-3458-2275]{Stephane Charlot}
\affiliation{Sorbonne Universit\'e, CNRS, UMR 7095, Institut d'Astrophysique de Paris, 98 bis bd Arago, 75014 Paris, France}

\author[0000-0002-6971-5755]{Gustavo Bruzual}
\affiliation{Instituto de Radioastronomía y Astrofísica, UNAM Campus Morelia, Apartado postal 3-72, 58090 Morelia, Michoacán, México}

\author[0000-0002-5189-8004]{Daniela Calzetti}
\affiliation{Department of Astronomy, University of Massachusetts Amherst, 710 North Pleasant Street, Amherst, MA 01003, USA}

\author[0000-0003-3667-574X]{You-Hua Chu}
\affiliation{Institute of Astronomy and Astrophysics, Academia Sinica No. 1, Sec. 4, Roosevelt Road, Taipei 10617, Taiwan}

\author[0000-0001-8587-218X]{Matthew J. Hayes}
\affiliation{Department of Astronomy and Oskar Klein Centre for Cosmoparticle Physics, AlbaNova University Centre, Stockholm University, SE-10691, Stockholm, Sweden}

\author[0000-0003-4372-2006]{Bethan L. James}
\affiliation{AURA for ESA, Space Telescope Science Institute, 3700 San Martin Drive, Baltimore, MD 21218, USA}

\author[0000-0002-6790-5125]{Anne. E. Jaskot}
\affiliation{Department of Astronomy, Williams College, Williamstown, MA 01267, USA}

\author[0000-0002-3005-1349]{G\"oran \"Ostlin}
\affiliation{Department of Astronomy and Oskar Klein Centre for Cosmoparticle Physics, AlbaNova University Centre, Stockholm University, SE-10691, Stockholm, Sweden}

\begin{abstract}
\noindent 
Mrk 71 is a low metallicity ($Z=0.16$~Z$_\sun$) starburst region in the local dwarf galaxy NGC 2366, hosting two super star clusters (SSCs A and B), and is recognized as a 
Green Pea (GP) analog with SSC A responsible for the GP properties. We present STIS and FOS far-ultraviolet (FUV) spectra of the embedded SSC \mka\ obtained with the Hubble Space Telescope (HST). The STIS FUV spectrum shows the characteristic features of very massive stars (VMS, masses $>100$~\Msun) and we derive an age of $1\pm1$~Myr by comparison with the Charlot \& Bruzual suite of spectral population synthesis models with upper mass limits of 300 and 600~\Msun.  We compare the STIS spectrum with all known SSC spectra exhibiting VMS signatures: NGC 5253-5, R136a, NGC 3125-A1 and the $z=2.37$ Sunburst cluster.
We find that the cluster mass-loss rates and wind velocities, as characterized by the C\four\ P Cygni profiles and the He\two\ emission line strengths, are very similar over  $Z=0.16$ to 0.4~Z$_\sun$. 
This agrees with predictions that the optically thick winds of VMS will be enhanced near the Eddington limit and show little metallicity dependence. We find very strong damped Lyman-$\alpha$ absorption with $N$(H\one) $=10^{22.2}$~\cm2\  associated with \mka. We discuss the natal environment of this young SSC in terms of radiatively-driven winds, catastrophic cooling and recent models where the cluster is surrounded by highly pressurized clouds with large neutral columns.
\end{abstract}


\section{Introduction} \label{intro}
The distinction between young massive star clusters (YMCs) found in nearby galaxies and globular clusters (GCs) formed at high redshift has become blurred in recent years, with the realization that both populations can be formed in intense star formation episodes when gas pressures are high \citep{elmegreen97, kruijssen15, elmegreen18}.  Super star clusters (SSCs, mass $> 10^5$~\Msun, radius $\la 1$~pc) are the most massive subset of YMCs and are only found locally in galaxy mergers, starburst dwarf galaxies and the centers of large galaxies. 

Studies of these GC-like systems are critical because they form under high pressure conditions that are similar to those found in star-forming galaxies at the peak of cosmic star formation. 
The strongly lensed SSC in the Sunburst Arc galaxy at $z=2.37$  \citep{rivera17b, chisholm19, vanzella20, vanzella22, mestric23}, and the cluster population of the lensed Sparkler galaxy at $z=1.38$ \citep{mowla22, claey23, adamo23} offer a rare insight into cluster formation at this peak epoch. 

It is thus essential to study local examples of young SSCs to investigate their massive star populations via FUV spectroscopy, and compare them to stellar population synthesis models, as a means of testing low metallicity models at the youngest ages representative of high-z systems, accessible with JWST spectroscopy.

Mrk~71 is a well-studied, low metallicity, local starburst region in the dwarf galaxy NGC~2366 at a distance of 3.44~Mpc \citep {tolstoy95} with $12+{\rm log(O/H)}=7.89$ 
\citep {izotov97,chen23} or 0.16 \Zsol\ using the solar oxygen abundance of \citet{asplund09}. 

\citet{micheva17} suggest that Mrk~71 is a local Green Pea (GP) analog. GPs are low metallicity, intensely star-forming galaxies with extremely strong [O\three] $\lambda5007$ emission and a typical redshift of $\sim 0.2$ due to the selection technique, which puts the strong [O\three] emission into the green band at this redshift (Cardamone et al. 2009). Mrk~71 shares many properties with GPs including a high ionization parameter, and may be  a good candidate for the escape of Lyman continuum photons \citep{komarova21}. The only substantial difference is that Mrk~71 is 1--2 orders of magnitude less luminous than GPs.

The starburst region of NGC~2366 contains two clusters \mka\ and \mkb\ whose properties were first investigated in the FUV by \citet{drissen00} with HST using the Faint Object Spectrograph (FOS). For \mka, they detected C\four\ $\lambda1550$ nebular emission and no underlying stellar population. \citet{drissen00} concluded that \mka\ is $\leq 1$~Myr old, in the massive ultracompact H\two\ region phase with the cluster embedded in natal molecular material, hidden from view in the FUV by dust, and is responsible for the bulk of the ionizing photons in the starburst.  They found that \mkb\ is older at 3--5~Myr and has evacuated its surrounding region. The two clusters have a projected separation of 5 arcsec or 83~pc. \citet{micheva17} show that \mka\ is responsible for the GP-like properties of Mrk~71. They derive a mass of $1.4\times 10^5$~\Msun\ by modeling the \ha\ luminosity.

Mrk~71 is also well known for having very broad emission wings with a full width zero intensity of $\ga 6000$~\kms\ \citep {roy92, g-d94, binette09} associated with the strong nebular lines of e.g., [O\three] and H$\alpha$.
\citet{komarova21} suggest that these features centered on \mka\ originate in a dense, clumpy, Lyman continuum or Lyman-$\alpha$-driven superwind. \citet{oey17} present CO($J=2-1$) observations of \mka\ and detect a compact ($\sim 7$~pc) molecular cloud with a mass of $10^5$~\Msun, which is similar to the mass of the SSC itself and implies a high star formation efficiency of 50\%. These observations suggest that stellar winds are not effective in fully clearing the natal gas at an age of 1~Myr and may be suppressed by catastrophic cooling due to the high densities in young compact SSCs \citep[eg.][] {silich04, silich17, silich18}. Radiation feedback should therefore dominate at the earliest ages as shown by \citet{komarova21} and predicted by \citet{silich13}. 
To further understand the role of suppressed winds, it is necessary to identify the massive star population of \mka\ to quantify the stellar winds, ionizing spectrum and verify its young age. 

The  well-studied young SSC NGC~5253-5 at the starburst center of the blue compact galaxy NGC~5253 has a similar age to \mka\ of 1~Myr \citep{calzetti15, smith16}. 
NGC~5253-5 was thought to have an age of 3--5~Myr from the presence of broad Wolf-Rayet emission lines in optical spectra \citep{monreal10} but \citet{smith16} showed that the stellar emission features arise from very massive stars (VMS, mass $> 100$~\Msun) at an age of 1--2~Myr by analyzing HST FUV spectroscopy obtained with the Space Telescope Imaging Spectrograph (STIS). 
NGC~5253-5 coincides with the peak of the \ha\ emission in the galaxy and \citet{smith16} showed that the high ionizing flux can only be explained by the presence of VMS. The presence of VMS in \mka\  has been suggested by \citet{james16} and \citet{micheva17} from the detection of He\two\ $\lambda4686$ emission in narrow-band imaging with the Wide Field Camera 3 (WFC3) on HST.

In Section~\ref{obs}, we present STIS FUV spectroscopic observations of \mka\ and show that VMS are present in the cluster in Section~\ref{spectra}. We examine the interstellar extinction and compare the spectrum with stellar population synthesis models in Section~\ref{models}. We compare the FUV spectrum of \mka\ with all known examples of SSCs containing VMS, and discuss what the STIS observations reveal about the local environment of \mka\  in Section~\ref{disc}. The summary and conclusions are presented in Section~\ref{summary} .

\begin{figure}
\includegraphics[width=0.45\textwidth,clip=true]
{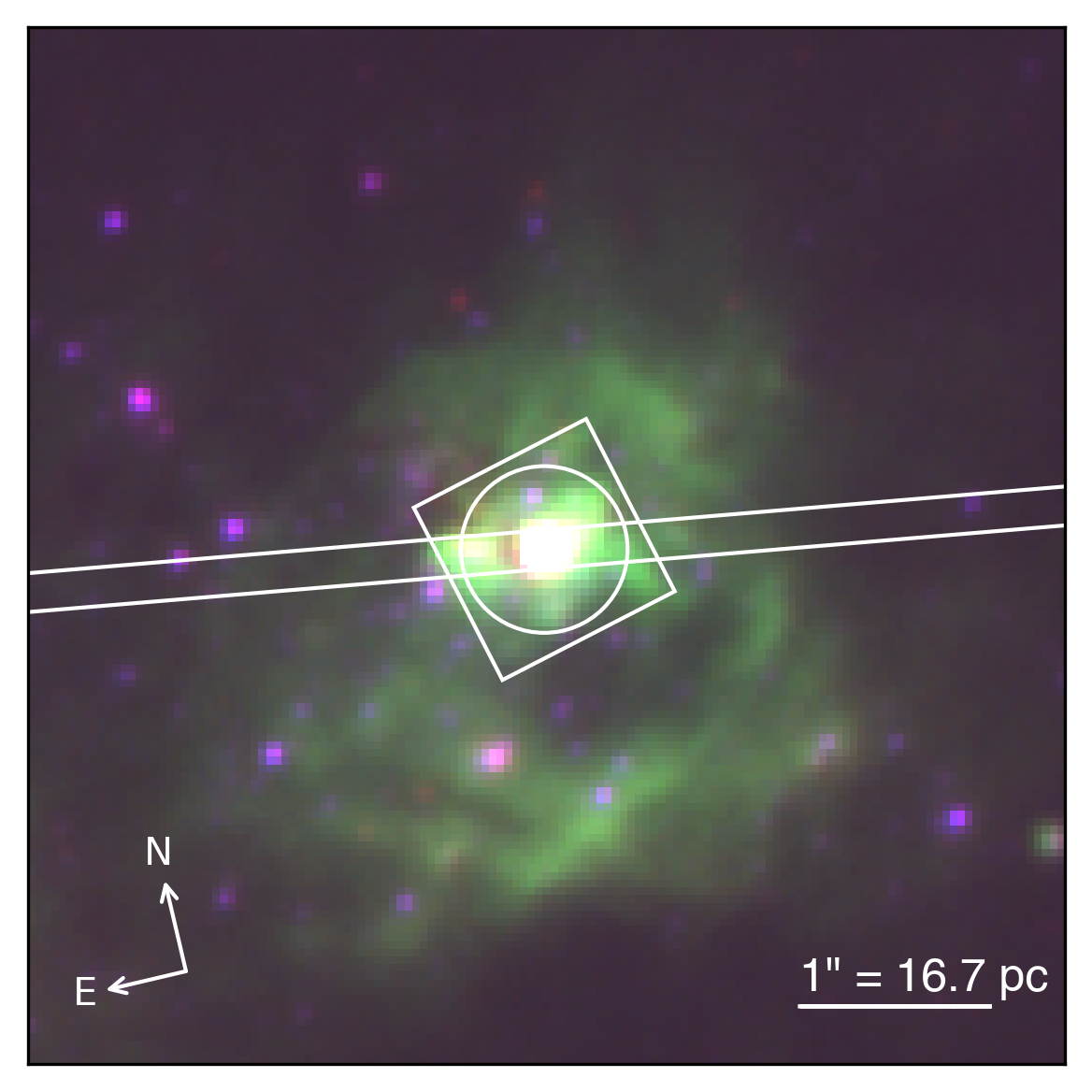}
\caption{WFC3/UVIS F438W (blue), F502N (green), F814W (red) image of \mka\ showing the positions of the long STIS slit and the FOS apertures (square
represents G190H and circle represents G130H observations).
}
\label{fig-STIS-slit}
\end{figure}
\section{Observations and Data Reduction}\label{obs}
We have obtained FUV long-slit spectroscopy with STIS of Mrk~71 as part of a Cycle 28 program (ID 16261; PI Oey) aimed at investigating the low metallicity feedback and stellar populations of SSCs \mka\ and -B via FUV imaging and spectroscopy. 
The FUV imaging results are reported in \citet{oey23}.
Here we report on the FUV spectrum of \mka. We supplement this dataset with archival Faint Object Spectrograph (FOS) observations of \mka\ (ID 6096; PI Drissen and ID 5246; PI Skillman).  In Fig.~\ref{fig-STIS-slit} we show the positions of the STIS and FOS apertures superimposed on WFC3/UVIS (ID 13041; PI James) images of \mka. 

\subsection {STIS Spectra}\label{stis-spectra}
Long-slit FUV spectra of \mka\ were obtained with HST/STIS in 2021 March and 2022 February. The dataset comprises 11 exposures over 4 visits (totaling 8.7 hr) taken with the G140L grating (1162--1712~\AA) and the $52''\times0''.2$ slit. The position angle was set at 81\degr.5 to obtain spectra of both \mka\ and -B. The image scale is $0.0246$ arcsec/pixel ($=0.41$ pc/pixel) and 0.58 \AA/pixel. Each observation was dithered along the slit using a 3-point dither pattern and a spacing of 12 pixels. 
\begin{figure*}
\centering
\includegraphics[width=18cm,clip=true]
{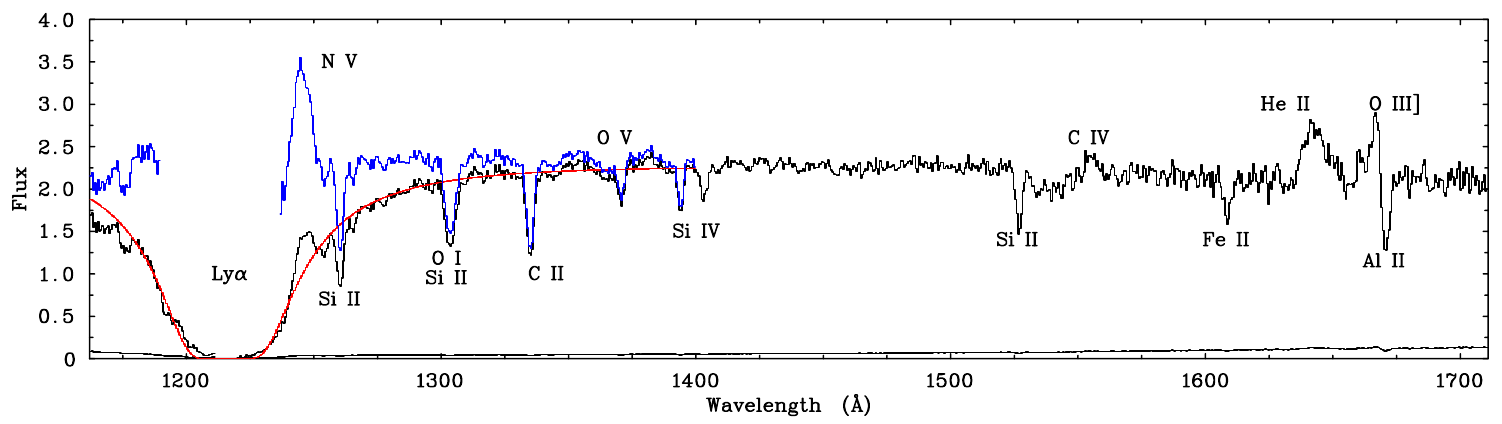}
\end{figure*}
\begin{figure*}
\centering
\includegraphics[width=18cm,clip=true]
{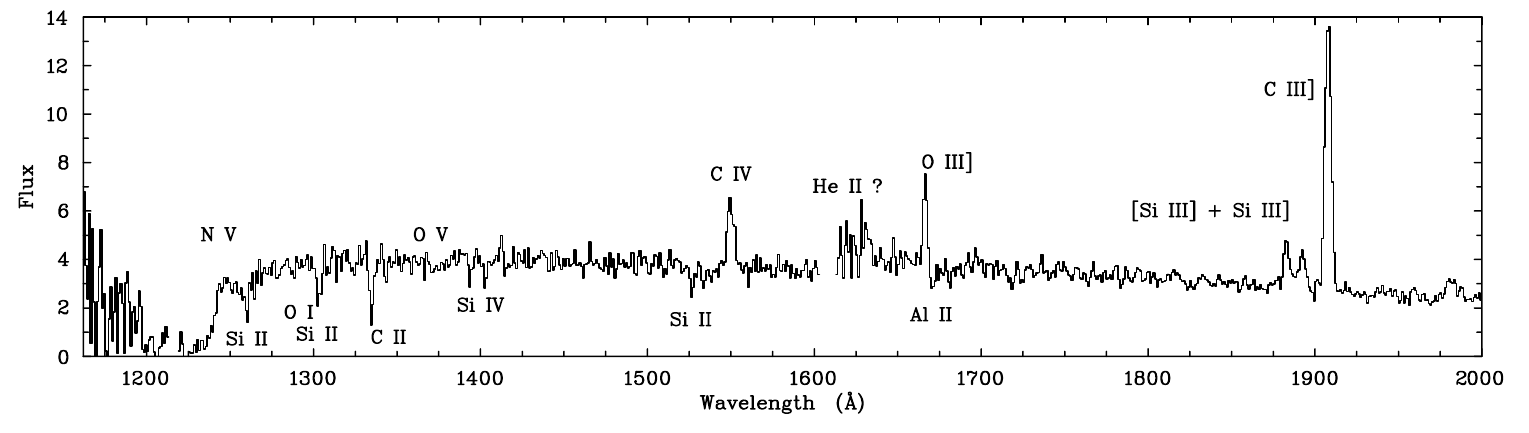}
\caption{Top panel: STIS G140L spectrum of \mka\ with the error array plotted along the bottom. The main stellar features and interstellar lines are identified. The damped Lyman-$\alpha$ fit is shown in red and the Ly$\alpha$-corrected N V emission line profile in blue.
Bottom panel: FOS G130H and G190H spectra of \mka. The main stellar features, interstellar and nebular lines are identified. The flux is in units of $10^{-15}$\, erg\,cm$^{-2}$\,s$^{-1}$\,\AA$^{-1}$.
}
\label{fig-spectra}
\end{figure*}

The individual 2D spectra were combined as follows. The {\sc stistools} \citep{hack18} package {\sc sshift} was used to align the flat-fielded (FLT) images for each visit after the dither offsets were verified by cross-correlating the brightest regions of the spectra. The aligned FLTs were then combined by averaging and weighting by the visit exposure
times. The error arrays were combined in quadrature as variance arrays with the square of the exposure times
as weights.  A flux-calibrated and distortion-corrected 2D spectrum was then created from the
combined FLT with the {\sc stistools} package {\sc x2d}. We selected a background region 340 pixels (=8\farcs4) below the SSC-A spectrum and 135  pixels (=3\farcs3) wide. In Fig.~\ref{fig-STIS-slit}, this region is to the west of \mka, outside of the field of view shown, and thus well away from the nebula.
The background region was averaged and fit with a fifth-order polynomial and then subtracted row by row from the combined X2D image. Finally, a 1D spectrum was extracted by summing over 11 pixels (=0\farcs27) in the spatial direction to optimize the signal-to-noise. 

The FWHM of \mka\ in the spatial direction along the STIS slit is measured to be $4.80\pm0.15$ pixels by fitting a Gaussian to the profile. From the STIS Instrument Handbook 
\citep{stisihb}, the line spread function (LSF) of a point source at 1500~\AA\ has a FWHM of 1.5 pixels for the same observational setup and thus \mka\ is resolved. Subtracting the LSF from the measured FWHM in quadrature, we derive a size of 109 mas or $0.93\pm0.03$~pc in radius. 
We also measured the FWHM in the spatial direction of an individual exposure to verify that the co-addition of the 11 spectra did not degrade the resolution;  we find a FWHM of $4.68\pm0.17$ pixels, giving a deconvolved radius of 0.91~pc, which is within the errors.

The deconvolved FWHM of \mka\ then gives a spectral resolution of 2.65~\AA\ or 550~\kms\ at 1450~\AA.
We confirm this value by fitting Gaussians to unresolved features (nebular O\three] $\lambda1661$, interstellar Si\two\ $\lambda\lambda 1526, 1260$) and find a mean FWHM of $2.60\pm0.16$~\AA. We correct for a radial velocity of $+95$~\kms\ \citep{micheva19} for Mrk~71 and leave the data in the original bin size of 0.58~\AA.

The signal-to-noise ratio of the final spectrum reaches a maximum of 46 at 1300~\AA\ and decreases to 20 beyond 1600~\AA\ where the G140L sensitivity declines. We present and discuss the spectrum in the next section. As can be seen in Fig.~\ref{fig-STIS-slit}, there is another object $0''.4$ to the E of \mka. The spectrum of this region is nebular and shows strong C\four\ and O\three]\ emission. The analysis of the nebular spectra and FUV imaging for \mka\ and -B will be presented in a separate paper.

\subsection{FOS Spectra}
\mka\  was observed with FOS in 1996 using the G130H grating (1140--1603~\AA) and the 0\farcs86 circular aperture with a total exposure time of 6150~s. The spectrum is presented and discussed in \citet{drissen00}. A longer wavelength spectrum was obtained in 1993 using the G190H grating (1603--2310~\AA) and the 1\farcs0 square aperture with a total exposure time of 3300~s. The spectrum covering C\three] $\lambda\lambda 1906,1909$ is discussed in \citet{garnett95} and \citet{rigby15}. We assume that the FOS spectral resolution is 3.2~\AA, as appropriate for an extended source, and use the re-calibrated spectra from the atlas of \citet{leitherer11}. We merged the two wavelength regions by using a scaling factor of 0.90 for the G130H spectrum. This value was derived by comparing the two FOS spectra with the STIS spectrum in the overlap region. 

\begin{deluxetable*}{lllrccccl}
\tablecolumns{8}
\tabletypesize{\footnotesize}
\tablecaption{Emission and Absorption Line Measurements for \mka.\label{tab1}}
\tablehead{
\colhead{Ion} & \colhead{Wavelength} & Line Type & \colhead{Velocity} & \colhead{FWHM} & \colhead{EW} & \colhead{$10^{15}$\,Flux} & \colhead{Instrument}
\\
\colhead{} & \colhead{\AA} && \colhead{\kms} & \colhead {\kms} & \colhead{\AA} & \colhead{erg\,s$^{-1}$\,cm$^{-2}$} 
\\
}
\startdata
N\five\ & 1238.82, 1242.80  & P Cyg em. & $+$560 & \nodata & $-3.9\pm0.1$ & \nodata & STIS\\
O\five\ & 1371.30 & Stellar abs.& $-165$ & \nodata & $+0.7\pm0.1$ & \nodata & STIS\\
&& P Cygni em. & $+1920:$ & \nodata & $-0.3\pm0.1$ & \nodata & STIS\\
&& P Cygni abs. & $-2765$ & \nodata & $+1.1\pm0.1$ & \nodata & STIS\\
C\four\ & 1548.20, 1550.78 & P Cyg em. & $+930$ & \nodata & $-0.5\pm0.1$ & \nodata & STIS \\
&& P Cygni abs.& $-3300$ & \nodata & $+1.5\pm0.2$ & \nodata & STIS\\
He\two\ & 1640.42 & Stellar em.  & $+450$ & $1770\pm150$  & $-3.0\pm0.2$ & $6.5\pm0.4$ & STIS\\
O\three] & 1660.81& Neb. em. & $+65$ &  $550\pm210$ &$-0.5:$ &1.1: & STIS\\
O\three] & 1666.15 & Neb. em. &$+40$& $550\pm210$ & $-1.3:$ & 2.7:& STIS\\
Si\two & 1260.42 & IS abs. & $+16$ & \nodata & $1.5\pm0.1$ & \nodata & STIS\\
O\one, Si\two & 1302.17, 1304.37 &  IS abs. & $+16$ & \nodata & $1.7\pm0.1$ & \nodata & STIS\\
C\two & 1334.53, 1335.71 &  IS abs. & $+35$ & \nodata & $1.6\pm0.1$ & \nodata & STIS\\
Si\four & 1393.75 &  IS abs. &  $+95$ & \nodata & $0.6\pm0.1$ & \nodata & STIS\\
Si\four & 1402.77 &  IS abs. &  $+60$ & \nodata & $0.6\pm0.1$ & \nodata & STIS\\
Si\two & 1526.71 &  IS abs. & $+18$ & \nodata & $0.9\pm0.1$ & \nodata & STIS\\
Fe\two & 1608.45 &  IS abs. & $-24$ & \nodata & $0.9\pm0.1$ & \nodata & STIS\\
Al\two & 1670.79 &  IS abs. & $+45$ & \nodata & $1.2\pm0.1$ & \nodata & STIS\\
C\four & 1549.49& Neb. em. & $+75$ &  \nodata & $-3.4\pm0.3$ & $14.0\pm1.2$ &FOS\\
O\three] & 1660.81,1666.15 & Neb. em. &$+50$& \nodata & $-3.3\pm0.2$ & 1$1.4\pm1.0$& FOS\\
$[$Si\three] & 1883.00 & Neb em. & $+10$ &  \nodata & $-3.2\pm0.2$ & $10.0\pm0.8$ &FOS\\
Si\three] & 1892.03 & Neb em. & $+25$ &  \nodata & $-2.7\pm0.2$ & $8.0\pm0.8$ &FOS\\
C\three] & 1907.71& Neb em. & $+10$ & \nodata & $-20.0\pm0.3$ & $53.6\pm2.1$ &FOS\\
\enddata
 \end{deluxetable*}

\section{Description of the Spectra}\label{spectra}
We show the STIS and FOS spectra of \mka\ in Fig.~\ref{fig-spectra} and label the main stellar features, nebular emission lines, and the strong interstellar (IS) absorption lines. Line measurements are provided in Table~\ref{tab1} where negative equivalent widths (EWs) indicate emission features.

One striking feature of the FUV spectrum is the strength of the damped Ly$\alpha$ absorption at 1215~\AA. The damping wings extend to $\sim 1400$~\AA\ and subsume the N\five\ P Cygni absorption profile. To derive the H\one\ column density and correct the N\five\ profile, the STIS spectrum was first normalized using a spline fit and extrapolated to the bluest wavelength available at 1162~\AA. Voigt profiles were then fitted to the spectrum using two components to represent the Milky Way (MW) and \mka\ Ly$\alpha$ absorption, following the approach of \citet{hernandez21}. For the MW component, a value of $\log N({\rm H\,I})=20.79~\cm2$ was adopted from the H\one\ map of \citet {HI4PI16}.  Radial velocities of $0$ (MW) and $+95$~\kms\  for \mka\ \citep{micheva19} were assumed and the routine {\sc voigtfit} \citep {krogager18} was used to fit the observed profile. 

We derive $\log N({\rm H\,I})=22.222\pm0.005$~\cm2 for the \mka\ line of sight. We discuss this high value in more detail in Section~\ref{ext} but note here that the high column of neutral hydrogen is indicative of a large amount of neutral gas associated with \mka. The resulting fit to the damped Ly$\alpha$ profile and the corrected N\five\ profile are shown in Fig.~\ref{fig-spectra}. We note that we cannot recover the absorption component of the N\five\  P Cygni profile because it has disappeared into the Ly$\alpha$ trough. We also note that the STIS spectral resolution at 1215~\AA\ of 650~\kms\ is insufficient to resolve any Ly$\alpha$ emission that may be associated with \mka\ from the geo-coronal component given the low radial velocity of $+95$~\kms.

The most conspicuous stellar features in the STIS spectrum are the strong N\five\  emission (once corrected for the damped Ly$\alpha$ absorption)
and the strong and broad (FWHM $=1770$~\kms) He\two\ $\lambda1640$ emission. 
O\five\ $\lambda1371$ is clearly detected and consists of blue-shifted stellar wind absorption and a possible weak, ill-defined P Cygni emission component.
This feature is rarely observed in cluster spectra because it is usually produced by the hottest and most massive O stars with short lifetimes.
Si\four\ P Cygni emission is absent, which is consistent with the other spectral features, indicative of the hottest and most massive stars where Si$^{3+}$ is photoionized to Si$^{4+}$ \citep{drew89}. The stellar wind column density of Si\four\ will also be low given the metallicity of Mrk~71 of $0.16 Z_\sun$.
The resonance doublet of C\four\ at $\lambda\lambda1548,1551$ is seen as a P Cygni profile and its weakness again demonstrates the low metallicity of the region.

These stellar diagnostic lines (O\five, N\five, He\two\ present and Si\four\ absent) together with the fact that \mka\ is clearly still embedded in its natal gas \citep[Fig.~\ref{fig-STIS-slit},][]{drissen00} indicate that \mka\ is very young and contains very massive stars.
\citet {crowther16}  used spatially resolved STIS FUV spectra of the R136a star cluster in 30 Doradus (age of 1--2~Myr) to show that the broad He\two\  
$\lambda1640$ emission is totally dominated by VMS in the integrated cluster spectrum at this young age.  Thus the  presence of broad He\two\ emission in \mka\ at a cluster age of $\sim 1$\ Myr \citep{drissen00} is a clear indicator of main sequence stars more massive than 100~\Msun\ (VMS). Likewise, such a strong N\five\ emission feature is only seen at the youngest ages. 

An alternative possibility is that the cluster is older ($>2.5$~Myr) and the spectral features are due to the presence of classical WR stars. WN stars would then produce the N\five\ and He\two\ emission but the standard WN N\four] $\lambda1486$ emission line is absent in the STIS spectrum (Fig.~\ref{fig-spectra}). In addition, WC stars are needed to account for the O\five\ feature but the weakness of C\four\ argues against their presence because the abundance of C is expected to be enhanced through core He-burning. We thus discount that classical WR stars are responsible for the UV spectral features of \mka\ and conclude that the cluster contains VMS.
 In Sect.~\ref{age}, we compare the spectrum to stellar population synthesis (SPS) models incorporating VMS.

The FOS spectra were taken with 0\farcs86 and 1\farcs0 arcsec apertures (as shown in Fig.~\ref{fig-STIS-slit}) and are dominated by strong nebular emission lines arising from the ultracompact H\two\ region. Overall, the spectral features are reminiscent of low metallicity, high redshift star-forming galaxies \citep [e.g.][]{senchyna22, schaerer22}.
C\three] $\lambda1909$ is very strong with an equivalent width of 20~\AA\ \citep [see also][] {rigby15} and nebular C\four\ emission is present in the FOS spectrum. The stellar features of N\five\ and O\five, although weak, are seen to be present when compared to the STIS spectrum. The G190H
spectrum is very noisy near the start at 1603~\AA\ and it is not clear if He\two\ $\lambda1640$ stellar emission is present because the potential noisy emission feature (marked ``He\two\ ?'' in Fig.~\ref{fig-spectra}) has a peak wavelength at 1628~\AA, although the wavelength calibration near the starting wavelength may be uncertain.

The ratio of [Si\three] $\lambda1883$ to Si\three] $\lambda1892$ can be used as an electron density diagnostic. We use {\tt PyNeb} \citep{luridiana15}  to derive a value of $n_{\rm e} = 7400\pm3900$ cm$^{-3}$ adopting an electron temperature $T_{\rm e}$ of $15\,000$\ K \citep{g-d94,chen23}. This corresponds to a high thermal pressure $P/k$ of $\sim 2\times10^8$~cm$^{-3}$\,K. \citet{mingozzi22} show that electron densities derived from the Si\three] lines are insensitive to electron temperature.

The derived density is much higher than the value of 800 cm$^{-3}$ quoted by \citet{perez01} from the [Ar\four] line ratio measured in ground-based spectra, while this refers to the larger scale H\two\ region rather than the inner 1 arcsec or 17 pc, as observed with FOS. \citet{micheva19}  derive values from ground-based IFU data of  $T_{\rm e} = 13400$~K and $n_{\rm e} = 273$~cm$^{-3}$ for \mka\ from the [S\two] doublet. Likewise, \citet{chen23} derive a lower $n_{\rm e}$ value of $160\pm10$~~cm$^{-3}$ from the [O\two] $\lambda\lambda 3726/3729$ ratio for Mrk~71. 
The much larger $n_{\rm e}$ obtained from [Si\three] compared to values derived from [Ar\four] and other optical diagnostics is similar to trends in $n_{\rm e}$ measurements for local blue compact starbursts in the CLASSY sample obtained by \citet{mingozzi22}. These authors find that UV density diagnostic line ratios are overall $\sim 2$ dex higher than their optical counterparts.
We note that the thermal pressure $P/k$ would be much lower if the optical values of $n_{\rm e}$ are adopted.
 
\section{Comparison with Models}\label{models}
\subsection{Dust Attenuation}\label{ext}
We derive the interstellar extinction $A_V$ towards \mka\ by comparing the observed continuum energy distribution
of the combined STIS and FOS spectrum to the continuum energy distributions predicted by stellar population synthesis (SPS) models for various values of $A_V$. We use the single star Binary Population and Spectral Synthesis (BPASS) version 2.2.1 models \citep{eldridge17, stanway18} with a \citet{kroupa01} initial mass function, an upper mass cutoff of 300 \Msun\  and a metallicity of $Z=0.003$. 
Although we use Charlot \& Bruzual (C\&B) models to fit the stellar spectral features in Section~\ref{age}, we note that the continuum shape is identical in both BPASS and C\&B models.

We merged the G190H FOS (scaled by 0.514) and the STIS spectra to provide a long wavelength baseline for determining the extinction. The spectrum was de-reddened for the Milky Way extinction using a value of E(B$-$V) of 0.033~mag \citep{schlafly11} with the \citet {cardelli89} law, corrected for a radial velocity of 95~\kms, and binned at 1~\AA\ intervals to match the BPASS model spectra.

The nebular continuum arising from ionized gas makes a significant contribution to the total continuum flux of clusters with ages $< 5$~Myr \citep{reines10} and therefore needs to be added to the model spectra. 
We calculated the nebular continuum by following the analytic approach of \citet{leitherer99}, which assumes all stellar photons below 912~\AA\ are converted into free-free, free-bound and two-photon emission at longer wavelengths. The resulting nebular continuum shape is simply scaled by the hydrogen ionizing photon rate $Q$(H\one) before adding to the model stellar continua.

We derived the best fitting extinction value $A_V$ by reddening the model spectra for ages of 1, 2 and 3~Myr using the average SMC Bar extinction law of \citet{gordon03} and comparing to the STIS$+$FOS spectrum using a $\chi^2$ approach \citep{smith06, westmoquette14}. We normalized the observed and model spectra between 2100--2300~\AA\  (no 2200~\AA\ extinction bump is detected) to provide maximum leverage for determining the reddening from the FUV spectral slope, 
and determined $\chi^2$ over the relatively featureless wavelength range 1410--1515~\AA. We find $A_ V = 0.23\pm0.02$ or $E(B-V) = 0.084\pm0.007$~mag for ages of 1--3~Myr. For comparison, \citet{james16} find a range of $E(B-V)$ from 0.13--0.23~mag for Mrk 71 as a whole using an LMC extinction law. \citet{micheva19} find $E(B-V)=0.21$~mag for \mka\ from the nebular Balmer emission lines. \citet{chen23} use the Balmer lines to derive $E(B-V)=0.06\pm0.03$ mag for Mrk~71 overall. We note that these other determinations are based on nebular line ratios and can be higher than the reddening determined from stellar continua for low metallicity galaxies \citep{shivaei20}.

\begin{figure}
\centering
\includegraphics[width=8.5cm,clip=true]
{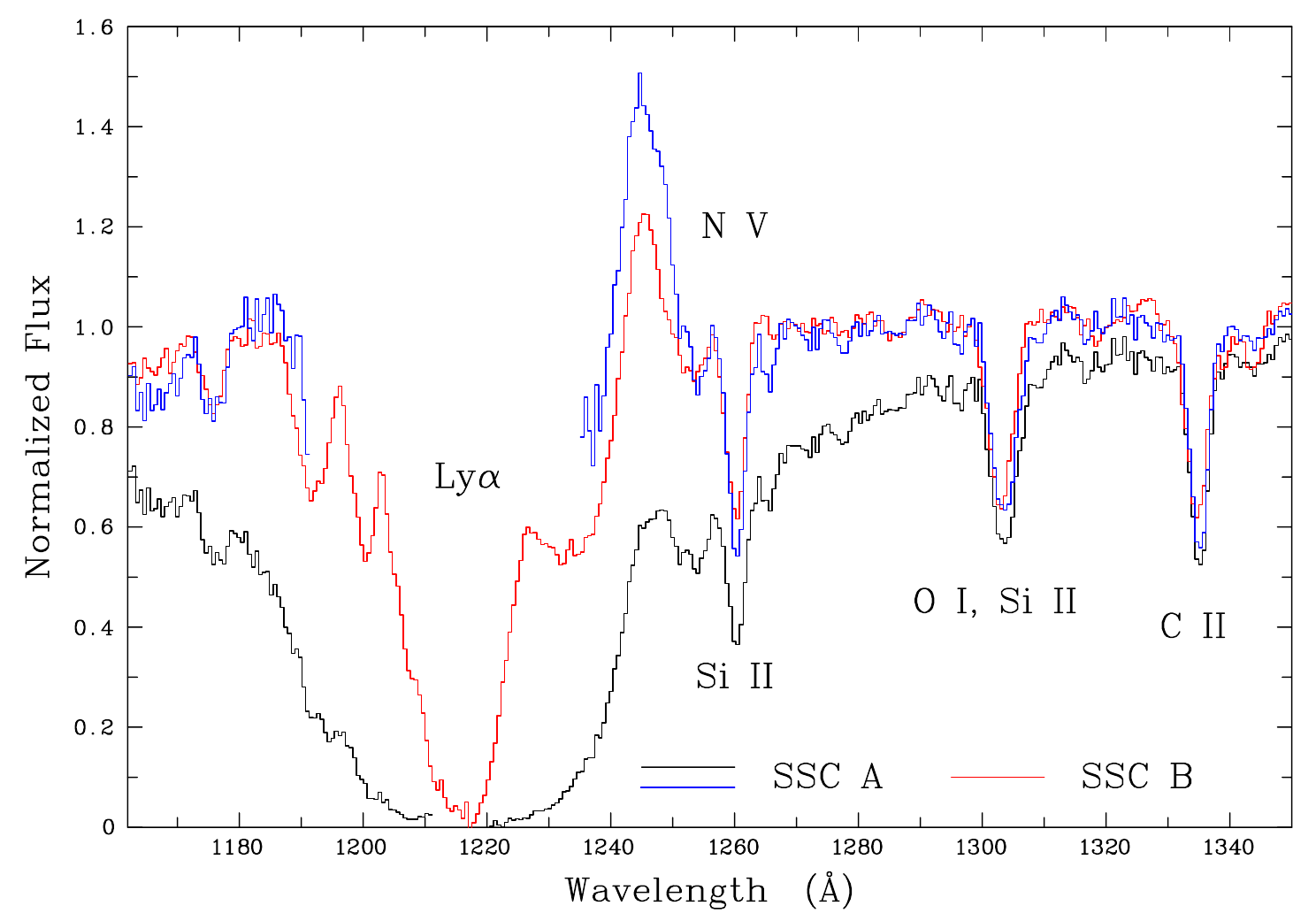}
\caption{Normalized STIS spectra of SSCs A and B in Mrk 71 in the region of the Ly$\alpha$ absorption feature and interstellar absorption lines of Si\two\ $\lambda1260$, O\one\ $\lambda1302$, Si\two\ $\lambda1304$ and C\two\ $\lambda\lambda$1334,1335. For SSC A, the observed spectrum (black)  and the Lyman-$\alpha$ corrected spectrum (blue) are shown. The observed spectrum of SSC A (black) should be compared with the spectrum of SSC B (red) to gauge the different strengths of the Ly$\alpha$ absorption. The spectrum for SSC A (blue) should be compared with the spectrum of SSC B (red) to gauge the similar strengths of the interstellar absorption lines.
}
\label{fig-IS}
\end{figure}

Overall, the derived extinction value is low considering the high column density of neutral hydrogen $\log N({\rm H\,I})=22.22$~\cm2 along the line of sight (Sect.~\ref{spectra}). The \citet{gordon03} SMC Bar relationship of $N$(H\one)/$A_ V=1.32 \times 10^{22}$~\cm2\,mag$^{-1}$
gives a much higher $A_V = 1.26$ or $E(B-V)= 0.46$ mag. If we scale this relationship to take account of the lower metallicity of \mka, we obtain 
$A_V = 0.81$~mag, which is still $\sim 3.5$ times higher than the derived value of 0.23 mag.
This implies there is little dust in the line of sight to \mka. 

We also have a STIS spectrum of the older, nearby cluster \mkb\ (projected separation of 83 pc) and
derive $\log N({\rm H\,I})=20.942\pm0.043$~\cm2 along its line of sight by fitting the Ly$\alpha$ absorption, taking into account a MW component, as was done for \mka\ in Sect.~\ref{spectra}. The Ly$\alpha$ absorption profiles for the two clusters are shown in Fig.~\ref{fig-IS}.
The H\one\ column density of \mka\ is a factor of 20 times higher than that observed along the line of sight of \mkb.
HST images of the Mrk 71 region show that \mkb\ has evacuated its surrounding gas and sits in a cavity \citep{james16}. The H\one\ column associated with \mkb\ therefore arises along the line of sight in the interstellar medium of the parent galaxy NGC~2366. The high H\one\ column density measured for \mka\ is clearly local to the embedded SSC and has a value of $\log N({\rm H\,I})=22.20$~\cm2 when corrected for the likely ISM component of NGC~2366.

Turning to the strengths of the interstellar metal absorption lines in the STIS spectra of \mka\  and \mkb, we also show the wavelength region covering the 
Si\two\ $\lambda1260$, O\one\ $\lambda1302$, Si\two\ $\lambda1304$ and C\two\ $\lambda\lambda$1334,1335 transitions in Fig.~\ref{fig-IS} for both clusters. We note that the IS absorptions are very similar in strength despite the factor of 20 difference in the $N$(H\one) column. 
\citet{micheva17} discuss the strengths of the Si\two\  IS absorption lines towards \mka\ using the measurements of \citet{leitherer11} taken from the FOS spectrum. They find a Si\two\ $\lambda 1260/\lambda 1526$ ratio of 6.0, indicating that the lines are optically thin. They indirectly derive an upper limit of $\log N({\rm H\,I}) < 20.0$~\cm2 and discuss this in the context of Lyman continuum escape. 

With the benefit of having a higher resolution and higher S/N STIS spectrum for \mka, we find that the Si\two\ $\lambda\lambda 1260,1526$ absorption lines have equivalent widths of 1.58 and 0.93~\AA\  respectively, giving a Si\two\ $\lambda 1260/\lambda 1526$ ratio of 1.7. This intermediate ratio between optically thick and thin gas, together with the fact that the residual intensities of the absorption lines are close to 0.6 (Fig.~\ref{fig-IS}),  suggests that the Si\two\  IS lines are optically thick but with a low gas covering fraction \citep{rivera17a,oestlin21}.

The overall similarity in the metal interstellar features along the lines of sight to \mka\ and \mkb, as shown in Fig.~\ref{fig-IS}, indicates that they are probably formed along the line of sight within the galaxy and are not local to the clusters, as expected for a low-ionization species.
It is puzzling that no distinct metal IS absorption features are associated with the high column of H\one\ gas in front of \mka. Higher resolution FUV spectra that separate out the MW and NGC 2366/Mrk~71 components are necessary to investigate this in more detail.

\begin{figure*}
\centering
\includegraphics[width=18cm,clip=true]
{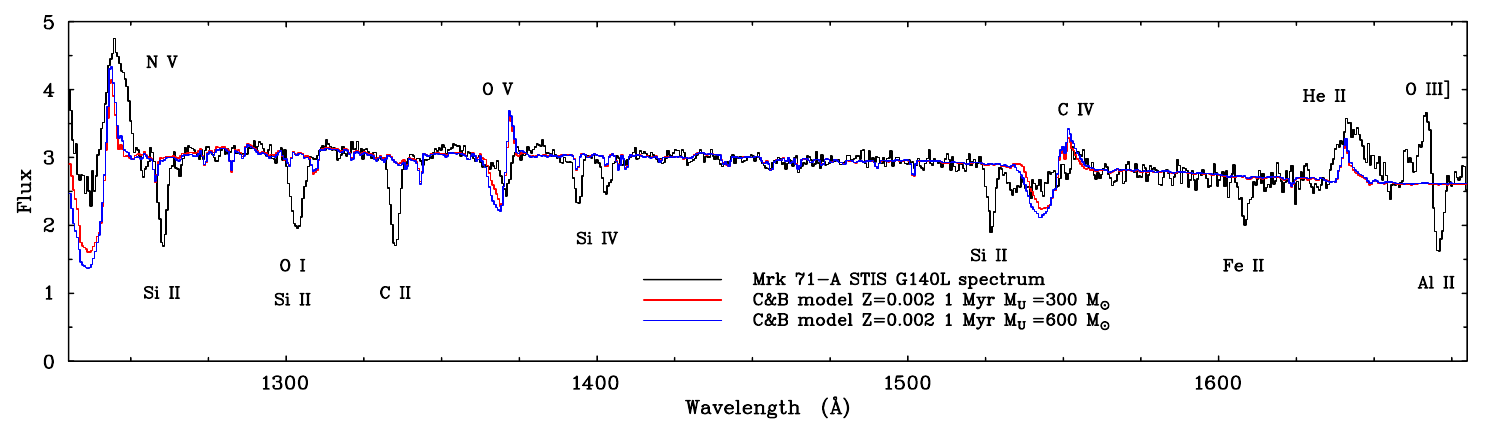}
\end{figure*}
\begin{figure*}
\centering
\includegraphics[width=18cm,clip=true]
{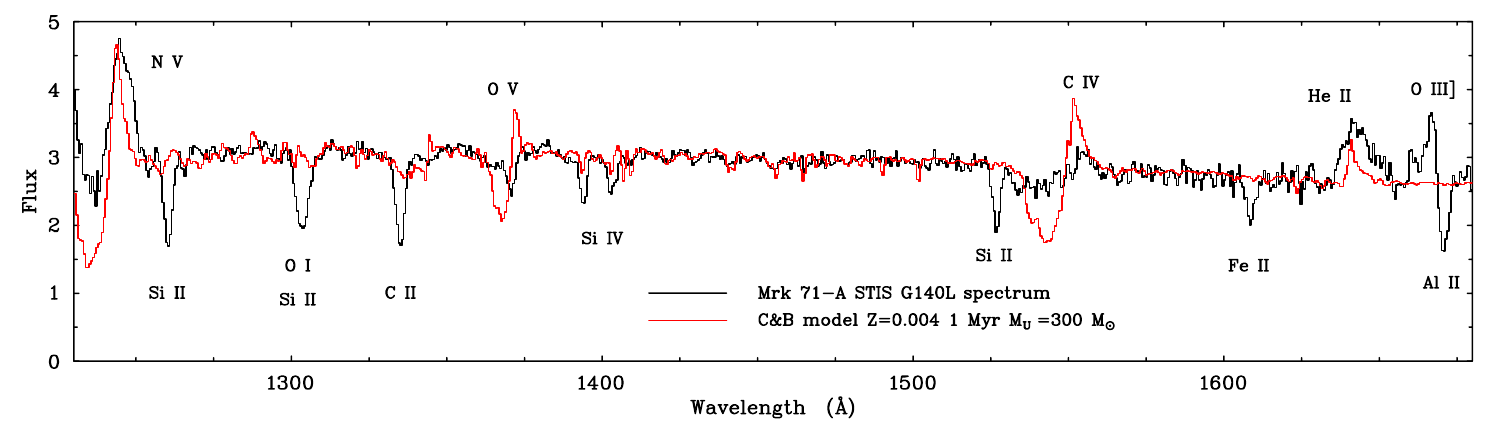}
\caption{Top panel: Comparison of the \mka\ STIS spectrum with the best-fitting C\&B models (upper mass limits $M_U=300, 600$ \Msun, $Z=0.002$, age $= 1$~Myr). The main stellar features (above the spectrum) and interstellar lines (below the spectrum) are identified. Bottom panel: Comparison for a C\&B model with $M_U=300$ \Msun, $Z=0.004$, age $= 1$~Myr.
The flux is in units of $10^{-15}$\, erg\,cm$^{-2}$\,s$^{-1}$\,\AA$^{-1}$.}
\label{fig-model-fit}
\end{figure*}

\subsection{Age and modeling of VMS features}\label{age}
A common technique to determine ages of young massive clusters is to model the UV spectral features using stellar population synthesis models \citep[e.g.][]{sirressi22}. It is well known, however, that SPS models with upper mass cutoffs of 100~\Msun\ cannot reproduce stellar He\two\ $\lambda1640$ emission at ages of $< 3$~Myr because of the absence of VMS in the models \citep[e.g.][]{wofford14, smith16, senchyna17,leitherer18}.
To overcome this deficiency, SPS models are required to incorporate VMS model atmospheres and evolutionary tracks that adopt realistic mass-loss rates for the optically thick winds of VMS. 

Recently, \citet{martins22} have produced the first SPS models including VMS that are tailored to LMC metallicity or 0.4~\Zsol. They adopt empirical mass-loss rates for VMS in R136a \citep{bestenlehner14} for optically thick stellar winds, and the standard mass-loss rates for optically thin O star winds derived by \citet{vink01}.  \citet{martins22} compare their FUV synthetic spectra at ages of 1 and 2~Myr with the integrated R136a spectrum \citep{crowther16} and NGC 3125-A1 \citep{wofford14} and find reasonable agreement. 

Overall, the work of \citet{martins22} shows that it is possible to match the strengths of the UV spectral features, particularly the He\two\ emission at LMC metallicity when the correct mass-loss rates are used to account for the optically thick winds of VMS. 

We compare the FUV spectrum of \mka\ with the updated SPS models from \citet{bruzual03}, which we refer to as Charlot and Bruzual or C\&B models for single stars. These models have the advantage that they cover metallicities down to $Z=0.0001$ and thus are more suitable than the \citet{martins22} models, which are tailored for LMC metallicity. 

The revisions to the \citet{bruzual03} models are described in \citet{plat19} and \citet{sanchez22}. They include
updated stellar evolutionary tracks from \citet{chen15} for masses up to 600~\Msun, which adopt the mass-loss formalism of \citet{vink11}, where
mass-loss rates are enhanced as the stellar luminosity approaches the Eddington limit, and the metallicity dependence decreases. This scheme allows for relatively higher mass-loss rates for VMS at low metallicity. The C\&B models utilize theoretical spectral libraries from \citet{leitherer10} and \citet{chen15} for O stars computed with {\sc wm-basic} \citep{pauldrach01} and the {\sc powr} library for WR stars \citep{hamann04}.

Recently, \citet{wofford23} have used the C\&B models to fit the HST/COS FUV spectrum of the SSC NGC~3125-A1, which has strong He\two\ $\lambda1640$ emission and O\five\ $\lambda1371$ absorption present, suggestive of VMS \citep{wofford14}. They find an excellent fit to the spectrum for $Z=0.008$ and an age of 2.2~Myr using an upper mass limit of 300~\Msun. Previous attempts at modeling the FUV spectrum with Starburst99 \citep{leitherer99} could not reproduce the strength of  He\two\ $\lambda1640$  without invoking a flat IMF exponent at an age of 3~Myr with an upper mass limit of 100~\Msun\ \citep{wofford14}.

\citet{senchyna21} compare the C\&B models with HST/COS FUV spectroscopy of 7 nearby star-forming regions at 0.2--0.4~\Zsol\ that exhibit broad 
(1500--2000~\kms) He\two\ $\lambda1640$ emission. They find that the models for continuous SF cannot simultaneously match the UV stellar wind lines and the optical nebular diagnostic lines. The model fits underestimate the strength of the He\two\ emission and C\four\ P Cygni absorption and emission, indicating a higher stellar metallicity is required, which is not supported by the nebular lines. To explain this mismatch, \citet{senchyna21} suggest that very massive stars formed through binary mass transfer and mergers unaccounted for in the models could explain the under-fitting of the stellar wind lines.

To compare the spectral features present in the STIS FUV spectrum of \mka\ we use the C\&B suite of SPS models for $Z=0.002$ and 0.004 with upper mass limits of 300 and 600~\Msun\ and a \citet{chabrier03} initial mass function. The nebular continuum, scaled by $Q$(H\one), is added to the model spectrum, and this is reddened by $A_V=0.23$~mag using the \citet{gordon03} extinction law. The model spectra are then binned to 0.584~\AA\ to replicate the STIS spectrum, which is corrected for foreground MW reddening, Ly$\alpha$ absorption, and radial velocity. The models are normalized to the STIS spectrum over the wavelength range 1420--1500~\AA.

\begin{figure*}
\centering
\includegraphics[width=18cm,clip=true]
{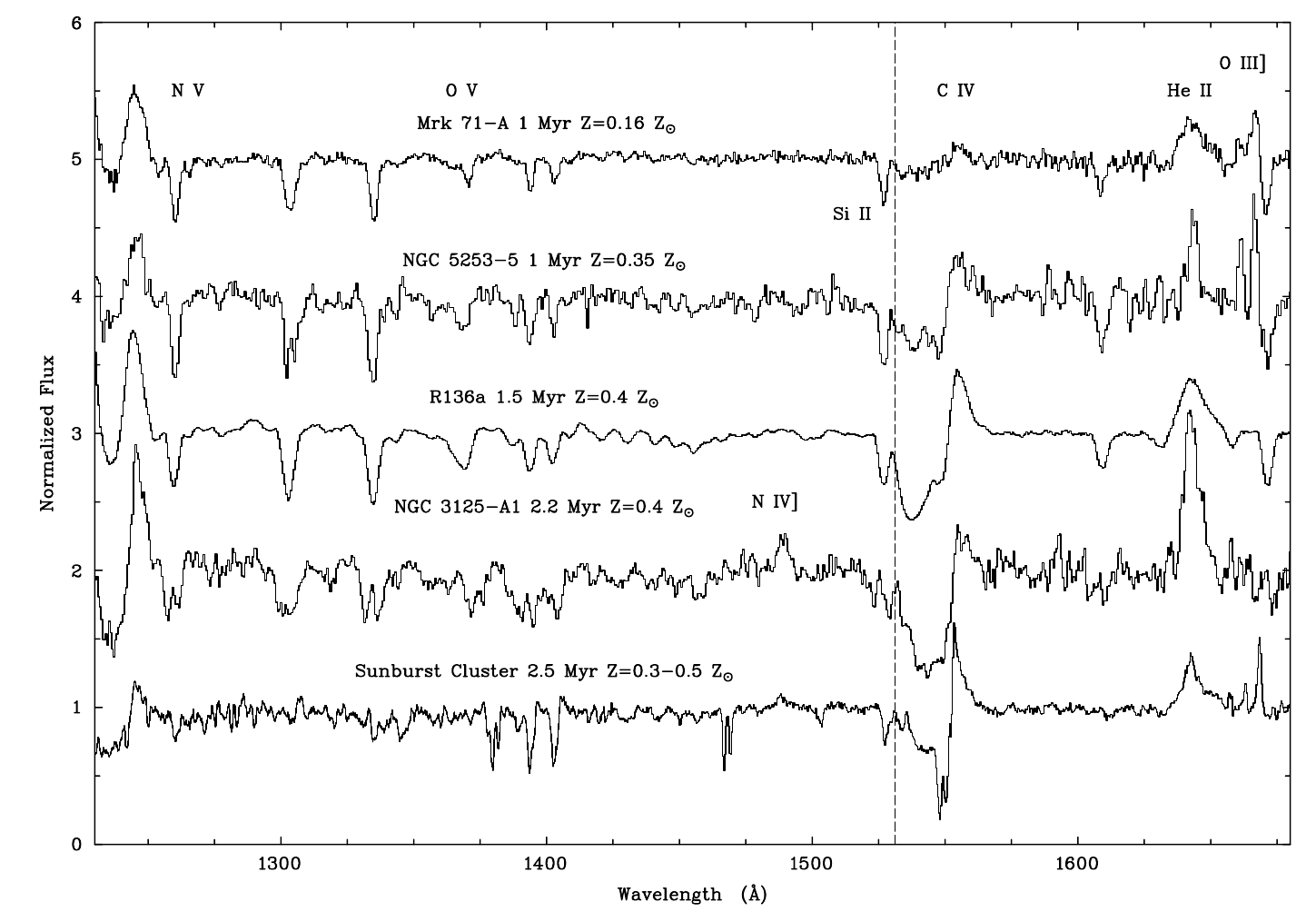}
\caption{Comparison of normalized FUV spectra of SSCs containing VMS. The spectra are ordered in increasing age and $Z$ from top to bottom. All the spectra, except for the Sunburst Cluster, were obtained with the STIS G140L grating. The vertical dashed line represents the maximum wind velocity of $-3300$~\kms\ measured for the C\four\ P Cygni absorption in \mka.
}
\label{fig-vms-spectra}
\end{figure*}

Comparison of the C\&B models with the observed \mka\ spectrum shows that O\five\ wind absorption is only predicted to be present for ages $< 2.5$~Myr \citep{wofford23} and thus we do not consider models with older ages. The He\two\ $\lambda1640$ emission strength is under-predicted in all the models we consider. The strongest He\two\ emission occurs for ages between 0.9--1.1~Myr and we show the model fit for 1~Myr in Fig.~\ref{fig-model-fit} for upper limits of 300 and 600~\Msun. There is very little difference between the predicted spectra for the two mass upper limits or the age range 0.9--1.1~Myr. 

The C\four\ P Cygni emission component is well-fitted but the predicted absorption component is too deep and the wind velocity is too low. 
We have considered whether the weakness of the C\four\ P Cygni absorption in the data could be due to dilution by a larger nebular continuum contribution, given the high surface brightness of the high density nebular gas (Section~\ref{spectra}). We increased the nebular continuum flux in the model by a factor of 2 at 1500~\AA\ (representing 75\% of the stellar flux)  and find that the depth of the residual absorption decreases from 0.79 to 0.84 of the continuum level, compared to the observed value of 0.92 for the C\four\ absorption in the \mka\ spectrum. Increasing the nebular contribution by this amount does dilute the C\four\ P Cygni absorption but not by the required amount. A larger nebular continuum contribution will also affect the UV continuum shape by increasing the continuum flux beyond 1500~\AA\ relative to shorter wavelengths. This effect is not seen in the observed spectra. 

The O\five\ wind absorption is well-matched in strength but there is a velocity offset and the emission component is too strong. Finally, N\five\ is too weak and narrow although the large damped Ly$\alpha$ correction renders this feature uncertain in the observations. The narrow widths of the C\four\ absorption and He\two\ emission indicate that the wind outflow velocities at $Z=0.002$ in the models are underestimated. 

We also compare the \mka\ STIS spectrum with C\&B models for $Z=0.004$ in the lower panel of Fig.~\ref{fig-model-fit}. The He\two\ emission is under-predicted again, as expected, and the C\four\ P Cygni emission and absorption are too strong, presumably because of the higher metallicity. In summary, the best fitting SPS model is for $Z=0.002$ and $1\pm1$~Myr. Models at these ages produce the strongest predicted He\two\ although the strength and width are not well-matched. The derived age is in excellent agreement with the value of $\le 1$~Myr estimated by \citet{drissen00}.

In summary, the C\&B models are not able to reproduce most of the stellar wind features in the \mka\ FUV spectrum. In particular,
the He\two\ flux and line width are both under-predicted. This is in contrast to the excellent fit to the spectrum of NGC 3125-A1 at LMC metallicity by \citet{wofford23} using the C\&B models. 
We note that the C\&B models are incomplete because they do not include tailored atmosphere models for VMS with optically thick winds. Instead they 
rely  on the {\sc wm-basic} O star atmosphere models to represent VMS at young ages and these have optically-thin winds. As shown by \citet{crowther16}, He\two\ $\lambda1640$ emission is exclusively produced by VMS at ages of 1--2~Myr, and this feature will be too weak and too narrow in SPS models that do not include optically thick winds for VMS. For NGC 3125-A1, \citet{wofford23} determine an age of 2.2~Myr and find that WR stars with masses $> 100$~\Msun\ start to appear at this early age in the models with an upper mass limit of 300~\Msun\ and $Z=0.008$.  Thus, the C\&B models switch to using optically thick wind models to account for these late WN (WNL) stars, and this will enhance the He\two\ emission line strength sufficiently to agree with the observations. WR stars are lacking at all ages in the Z=0.002 C\&B models.

\section{Discussion}\label{disc}
\subsection{Comparison with other SSC VMS Spectra}\label{disc1}

In Fig.~\ref{fig-vms-spectra} we compare the FUV spectrum of \mka\ with all known examples of VMS in SSCs. The spectra are ordered by increasing age and $Z$ from top to bottom. The STIS spectra of NGC~5253-5 \citep{smith16}, R136a \citep{crowther16} and NGC 3125-A1 \citep{wofford14} are shown together with the VLT MUSE and X-Shooter spectrum of the $z=2.37$ Sunburst cluster (source 5.1) \citep{mestric23}. The STIS spectra have been corrected for damped Ly$\alpha$ absorption, continuum-normalized to aid comparison, and aligned in wavelength using the He\two\ emission feature. 

The blue compact dwarf galaxy NGC~5253 hosts a central young starburst containing 3 SSCs \citep{smith20}. The cluster NGC~5253-5 coincides with the peak of the H$\alpha$ emission in the galaxy and is visible at FUV wavelengths. R136a is the resolved, central ionizing cluster of the 30 Doradus H\two\ region in the LMC and the STIS spectrum from \citet{crowther16} represents the integrated light of R136a. The cluster A1 in the blue compact dwarf galaxy NGC 3125 is well known for having very strong He\two\ $\lambda1640$ emission \citep{chandar04} that could not be modeled with SPS models assuming a Wolf-Rayet origin \citep{wofford14}. As described in the previous section, \citet{wofford23} have successfully fitted the FUV spectrum with C\&B models including VMS. The final object shown in Fig.~\ref{fig-vms-spectra} is the cluster (source 5.1) in the lensed Sunburst Arc galaxy at $z=2.37$ from \citet{mestric23} who identify VMS spectral features and fit the spectrum using the \citet{martins22} SPS models including VMS.  A metallicity of $Z=0.5~Z_\sun$ was derived by \citet{chisholm19} and $0.3~Z_\sun$ by \citet{pascale23}.

We now compare the VMS spectral features shown in Fig.~\ref{fig-vms-spectra} with the aim of providing insights on the cluster winds at young ages and as a function of metallicity. It is clear that \mka\ has the lowest metallicity because the C\four\ P Cygni absorption feature is very weak compared to the other clusters, which all have LMC-like metallicities. 
\citet{chisholm19} show that the C\four\ absorption strength is a good metallicity indicator using Starburst99 SPS models \citep{leitherer99}. This dependence is also shown in Fig.~\ref{fig-model-fit} for the C\&B models.

The blue edge of the C\four\  absorption profile is one of the main observational diagnostics of stellar wind velocities in O stars. 
The scaling of the wind terminal velocity $v_{\infty}$ with $Z$ is not well constrained from observations because of the difficulty of measuring stellar wind velocities at low $Z$ where the C\four\ absorption is weak and unsaturated. For SPS modeling as a function of $Z$, the relationship of \citet{leitherer92} derived from radiatively-driven wind theory is usually adopted with $v_\infty \propto Z^{0.13}$. 

We can examine the $Z$ dependence by comparing wind velocities in Fig.~\ref{fig-vms-spectra}. The dashed vertical line at $-3300$~\kms\ represents the measured maximum wind velocity $v_{\rm edge}$ for \mka. It can be seen that the local clusters at LMC-like metallicity have very similar wind velocities and thus there appears to be little if any scaling of wind velocity with $Z$.  The Sunburst Cluster at $z=2.37$ does, however, appear to have a lower maximum wind velocity of $-2300$~\kms.

Empirical measurements of wind velocities in the literature for individual O stars as a function of $Z$ show contrasting results. The large HST program Ultraviolet Legacy Library of Young Stars as Essential Standards (ULLYSES)\footnote{https://ullyses.stsci.edu/index.html} \citep{duval20}  is set to improve our understanding of OB stars as a function of $Z$ by the analysis of the UV spectra for a significant number of OB stars in Local Group galaxies. \citet{hawcroft23} used 149 OB stars in the LMC and SMC from the ULLYSES dataset to measure the dependence of the wind terminal velocity on $Z$. They find that $v_\infty \propto Z^{0.22}$, which is steeper than the theoretical prediction of $v_\infty \propto Z^{0.13}$ \citep{leitherer92}. The earlier study of \citet{garcia14} determined wind velocities for 8 OB stars in IC 1613 (Z$=0.13~Z_\sun$) and found no clear differences between IC 1613, SMC or LMC OB stars. ULLYSES should improve on this small sample by increasing the dataset to $\sim 30$ OB stars in Local Group galaxies below SMC metallicity.

From the above, and given that C\four\ absorption originates in O stars, we would expect to see a lower maximum wind velocity for \mka\ for its lower $Z$ but this is not seen and is instead similar to LMC cluster values. \citet{garcia14} suggest that the similar wind velocities they find for IC 1613 OB stars compared to similar stars in the LMC and SMC could be due to an enhanced Fe abundance in this galaxy. \citet{drissen01} discuss the Fe abundance in the Luminous Blue Variable star V1 in Mrk 71 and find that it is SMC-like from modeling the strengths of the Fe\two\ absorption lines.
We note, however, that Fe is the main driver of mass loss in the inner wind while C, N and O are responsible for the wind acceleration to terminal velocity in the supersonic regime \citep{vink22}. 

We now discuss the strength and width of the He\two\ $\lambda1640$ emission line,  which is found exclusively in VMS \citep{crowther16} at very young ages ($< 2$~Myr) and are good indicators of wind density and velocity for the He\two\ formation region. 
 
 The He\two\ emission line profiles for \mka, R136a and the Sunburst Cluster are remarkably similar in strength and width with EWs of  $-3.0$ (\mka; Table~\ref{tab1}), $-4.4$ \citep[R136a;][]{crowther16} and $-3.0$~\AA\ \citep[Sunburst Cluster;][] {mestric23}.  Likewise, the FWHMs are 1770, 1970 and 1610~\kms. These similarities suggest that despite the difference in metallicity and redshift, the SSC winds in the VMS phase have comparable mass loss rates and velocities or feedback efficiencies. This is in contrast to the weak C\four\ P Cygni absorption feature in \mka, which is dominated by O stars, suggesting a mass loss rate much lower than LMC metallicity although the C\four\ wind velocities are similar.
 The other two SSCs NGC~5253-5 and NGC~3125-A1 have stronger and narrower He\two\ emission features. In Section~\ref{age}, we noted that the C\&B models at 2.2~Myr for NGC~3125-A1 contain VMS and WNL stars with the presence of WNL stars probably enhancing the He\two\ emission line strength. We thus speculate that the stronger and narrower He\two\ emission feature shared by these two SSCs may be due to both VMS and WNL stars whereas the He\two\ emission line profiles for \mka, R136a and the Sunburst Cluster are produced by VMS only.
 We note that \citet{wofford23} rule out a nebular contribution to He\two\ in NGC~3125-A1.

The auroral O\three] $\lambda\lambda 1661,1666$ emission lines are present in the spectra of \mka, NGC~5253-5 and the Sunburst Cluster (Fig.~\ref{fig-vms-spectra}). Both \mka\ and NGC~5253-5 are immersed in ultracompact H\two\ regions whereas R136a  and NGC 3125-A1, which do not show O\three], have been cleared of natal gas. This suggests that the Sunburst Cluster may be in the ultracompact H\two\ region phase. 

The strongest emission feature in the FUV spectra of the local SSCs in Fig.~\ref{fig-vms-spectra} is N\five\ $\lambda1240$; the strength of this feature is only apparent when the damped Lyman-$\alpha$ absorption feature has been removed.  We note that the two oldest clusters NGC 3125-A1 and the Sunburst show N\four] $\lambda1486$ emission in agreement with the VMS SPS model predictions of \citet{martins22}. These models predict a strong nitrogen enrichment after 1.5 Myr that boosts the strength of N\four] $\lambda1486$. O\five\ $\lambda1371$ is present in the local SSCs and appears as a blue-shifted absorption component with little to no emission, in contrast to SPS models that predict strong emission (Fig.~\ref{fig-model-fit}).  O\five\ $\lambda1371$ is absent in the Sunburst cluster and \citet{mestric23} argue that this is due to the older age of the cluster.

Overall, the similar emission line strengths and wind velocities for the SSC spectra shown in Fig.~\ref{fig-model-fit} argue for a weak dependency on metallicity. We find that the wind velocity scaling with $Z$ is close to constant as predicted by \citet{leitherer92}. The wind mass-loss rates also show little if any scaling with $Z$ as evidenced by the similar He\two\ emission line profiles at $Z=0.16$ and $0.4~Z_\sun$. This can be explained by the fact that VMS are expected to be close to their Eddington limits and their mass-loss rates will be strongly enhanced and the metallicity dependence for mass loss will decrease \citep{vink11, chen15}. For comparison, the mass-loss rates of optically thin O star winds are predicted to scale as $Z^{0.7-0.8}$ \citep{leitherer92, vink01}. The lack of scaling of the cluster wind parameters suggests that the VMS feedback efficiency is largely independent of metallicity over the range investigated for clusters at ages $<3$~Myr. 

\subsection{The local environment of \mka}\label{disc2}
We now consider the local environment of \mka\ to provide an overall view of a young SSC  embedded in its natal gas at low $Z$. At the center is the 1~Myr old SSC A (Sect.~\ref{age}) with a radius of 0.9~pc (Sect.~\ref{stis-spectra}) and a
mass of $1.4 \times 10^5$~\Msun\ \citep{micheva17}. \mka\ is ionizing a giant H\two\ region with a density up to $\sim7400$~cm$^{-3}$ in the immediate vicinity of the SSC and has a thermal pressure $P/k$ of $\sim 2\times 10^8$~cm$^{-3}$\,K, derived from the UV [Si\three] lines (Sect.~\ref{spectra}).

Neutral hydrogen is detected along the line of sight and shown to be associated with \mka\ with a high column density of $N$(H\one) $= 10^{22.2}$~\cm2\ but little dust (Sect.~\ref{ext}).  \citet{oey17}  detect a compact CO cloud with a size of 7~pc and mass of $10^5$~\Msun\ coincident with \mka. 
The presence of high density neutral and molecular gas co-located with the SSC is consistent with the findings of \citet{oey23} 
who use FUV nebular C\four\ imaging of \mka\  to study the mechanical feedback.
They show that the observed diffuse C\four\ surface brightness and its spatial distribution for the SSC A environment are both consistent with model predictions that it is undergoing strong radiative cooling, and driving a momentum-conserving shell due to catastrophic cooling. 

Feedback is thus dominated by radiation from the SSC, including from our newly identified VMS. This supports the scenario obtained by \citet{komarova21}:
The giant molecular cloud represents the remnant gas out of which the young SSC formed.  It is being fragmented by radiative feedback from the cluster, forming the radiation-driven superwind.  
The nature of the Lyman-continuum driven wind implies that there must be optically thin channels through which the Lyman continuum photons can escape, and the covering factor of the high column density clouds will then be less than unity. 

\citet{pascale23} model the Sunburst cluster, its escaping Lyman continuum photons \citep{rivera19} and ionized nebula. They find that the cluster is surrounded by  highly pressurized, dense clouds ($n_{\rm e} \sim 10^5$cm$^{-3}$), which should have large neutral columns ($N$(H\one) $> 10 ^{22.5}$~\cm2) to survive rapid ejection by radiation pressure. The parameters we find for \mka\ bear similarities to this model, particularly our measured high H\one\ column density of $N$(H\one) $= 10^{22.2}$~\cm2 and $n_{\rm_e} = 7400\pm3900$~cm$^{-3}$.

\section{Summary and Conclusions}\label{summary}

We have presented STIS and FOS FUV spectra of the local, low metallicity GP analog \mka\ with the aims of identifying the massive star population, verifying the young age, investigating the properties of stellar winds at low $Z$ and studying the embedded natal gas associated with this SSC.

The FOS spectrum \citep{drissen00} shows that \mka\ is a rare example of a high excitation, local starburst region with nebular C\four\ and strong C\three] emission (EW$=20$~\AA; Table~ \ref{tab1}). We are able to uncover the stellar spectral features with our deep and higher resolution STIS spectrum and show that the presence of O\five\ $\lambda1371$ and broad He\two\ $\lambda1640$ emission with the absence of Si\four\ $\lambda1400$ P Cygni emission indicates that VMS are present in this very young cluster.

We compare the STIS spectrum of \mka\ with the Charlot \& Bruzual suite of SPS models for upper limits of 300 and 600~\Msun\ and $Z=0.002$ and 0.004. For $Z=0.002$, we find that the He\two\ emission line strength is under-predicted in all the models and is strongest for ages of 0.9--1.1 Myr. There is very little difference in the fits for upper mass limits of 300 or 600~\Msun\ (Fig.~\ref{fig-model-fit}). Overall, the He\two\ emission in the models is too weak and narrow because {\sc wm-basic} O star atmosphere models are adopted to represent VMS and these have optically-thin winds. 
The C\four\ P Cygni absorption is too deep and the wind velocity is too low for $Z=0.002$ whereas the 
$Z=0.004$ models provide a poorer fit to the C\four\ P Cygni feature because the metallicity is too high. We derive an age based on the C\&B models of $1\pm1$~Myr, which is in excellent agreement with that estimated by \citet{drissen00}.

We compare the low metallicity STIS spectrum of \mka\ with all known examples of SSCs containing VMS: NGC 5253-5 \citep{smith16}, R136a \citep{crowther16}, NGC 3125-A1 \citep{wofford14} and the $z=2.37$ Sunburst cluster \citep{mestric23}. The comparison spectra have LMC-like metallicity and it is clear that \mka\ has the lowest $Z$ because the C\four\ P Cygni absorption is weak in comparison. We examine the $Z$ dependence of the cluster wind velocity and find that there appears to be little, if any, scaling with $Z$, despite theoretical predictions \citep{leitherer92} and recent measurements \citep{hawcroft23}.

The stellar He\two\ $\lambda1640$ emission line profiles in \mka, R136a and the Sunburst cluster are very similar in terms of strength and width and indicate similar wind densities and velocities irrespective of $Z$. We conclude that the VMS winds over $Z=0.16$--$0.4~Z_\sun$ have comparable mass-loss rates and velocities or feedback efficiencies. This agrees with the predictions of \citet{vink11} that the mass-loss rates of the optically thick VMS winds will be enhanced close to the Eddington limit and the metallicity dependence will decrease. 

Although some SPS models now extend to upper mass limits of 300~\Msun\ or higher, they lack tailored model atmospheres for VMS with their high mass loss rates and decreased metallicity dependence. The only example to date is the LMC metallicity models of \citet{martins22}. More VMS models are clearly needed to realistically model JWST spectra of low metallicity star-forming galaxies at high redshift when VMS, if present, will dominate the stellar wind and ionizing feedback in young globular clusters.

The STIS spectrum of \mka\ shows an unusually strong damped Lyman-$\alpha$ absorption feature with $N$(H\one) $=10^{22.2}$~\cm2\ that is associated with the SSC natal gas. We suggest that the covering factor of the H\one\ must be less than one to allow the Lyman continuum photons to escape. The adiabatic cluster wind is expected to be  suppressed due to catastrophic cooling because of the high densities and instead the presence of a Lyman continuum-driven wind is observed \citep{oey17, komarova21}. 
The model of the ionized nebula associated with the Sunburst cluster put forward by \citet{pascale23}  in which the cluster is surrounded by highly pressurized clouds with large neutral columns has many similarities to the properties we can measure for \mka.
\\
\\
\noindent 
We thank the referee for their astute and constructive comments on the manuscript.
We thank Uros M{e\v{s}}tri{\'c} and Eros Vanzella for kindly providing us with their Sunburst cluster spectrum.
We thank Fabrice Martins for sharing his VMS models and Calum Hawcroft for useful discussions on O stars winds.
BLJ is thankful for support from the European Space Agency (ESA). 
M.H. is a Fellow of the Knut \& Alice Wallenberg Foundation.
This work made use of v2.2.1 of the Binary Population and Spectral Synthesis (BPASS) models as described in \citet{eldridge17} and \citet{stanway18}.
Based on observations made with the NASA/ESA Hubble Space Telescope, at the Space Telescope Science Institute, which is operated by the Association of Universities for Research in Astronomy, Inc., under NASA contract NAS5-26555. These observations are associated with program \#16261. Support for program \#16261 was provided by NASA through a grant from the Space Telescope Science Institute, which is operated by the Association of Universities for Research in Astronomy, Inc., under NASA contract NAS 5-26555.

The data presented in this paper can be obtained from the Mikulski Archive for Space Telescopes (MAST) at the Space Telescope Science Institute. The specific observations analyzed can be accessed via \dataset[10.17909/ye2e-af62]{https://doi.org/DOI}.

\facility{HST (STIS, FOS)}
\software{STISTOOLS, VOIGTFIT, PYNEB}
\bibliographystyle{aasjournal}
\bibliography{references}

\end{document}